\documentclass[twocolumn,amsmath,amssymb]{revtex4}
\usepackage[utf8x]{inputenc}
\usepackage{hyperref}
\usepackage{graphicx}
\usepackage{color}

\begin{document}

\title{Generalizing a Unified Model of Dark Matter, Dark Energy, and
Inflation with Non Canonical Kinetic Term}

\author{Josue De-Santiago}
\email{josue@ciencias.unam.mx}
\affiliation{Depto. de F\'{\i}sica, Instituto Nacional de Investigaciones
Nucleares, M\'{e}xico.}

\author{Jorge L. Cervantes-Cota}
\email{jorge.cervantes@inin.gob.mx}
\affiliation{Berkeley Center for Cosmological Physics, LBNL and 
University of California, Berkeley, California 94720, USA}
\affiliation{Depto. de F\'{\i}sica, Instituto Nacional de Investigaciones
Nucleares, M\'{e}xico.}

\date{\today}

\begin{abstract}
We study a unification model for dark energy, dark matter, and inflation with a single 
scalar field with non canonical kinetic term.   In this model 
the kinetic term of the Lagrangian accounts for the dark matter and dark 
energy,  and at early epochs a quadratic potential accounts for slow roll   
inflation. The present work is an extension to 
the work by Bose and Majumdar [Phys. Rev. D 79, 103517 (2009).] with a more general kinetic
term that was proposed by Chimento in Phys. Rev. D 69, 123517
(2004).  We demonstrate that the
model is viable at the background and linear perturbation levels. 
\end{abstract}
\maketitle

\section{Introduction}

In some recent works, the possibility to unify  
the description of dark matter, dark energy, and inflation it has been studied . For example, in the
work by Liddle and Ure\~na-Lopez \cite{Liddle:2006qz} is studied
through a single scalar field that in the early Universe can account for the
inflaton field and at late times oscillates around a nonzero minimum of the
potential to yield the dark matter and dark energy components  the unification. The key issue there 
is that the reheating epoch must reduce the field density enough to be
subdominant during the radiation epoch, but not completely, for it to 
account for the dark matter, something that proved to be nontrivial to achieve
in standard reheating schemes \cite{Kofman:1997yn}.  Further works tried to
solve the issue by considering plasma masses \cite{Cardenas:2007xh,
Panotopoulos:2007ri}, a brief thermal inflation period
\cite{Liddle:2008bm}, and braneworld cosmological equations of motion
\cite{Lin:2009ta}.
Another unification model was proposed in \cite{Capozziello:2005tf}, where 
a phantom scalar field accomplishes for a $\omega<-1$
period in inflation and in the late time Universe. In
\cite{Henriques:2009hq} unification is achieved from the 
compactification of the six dimensional supergravity Salam-Sezgin model, where  
an analysis is included on the gravitational waves  power spectrum.

On the other hand,  the idea that a modified kinetic term in the
Lagrangian could produce accelerated expansion was first proposed in the
context of inflation \cite{ArmendarizPicon:1999rj} and later in 
dark energy \cite{Chiba:1999ka}. This work was then generalized
in what now we call k-essence \cite{ArmendarizPicon:2000dh,
ArmendarizPicon:2000ah}. This kind of model has been
used in several works to unify dark energy and dark matter
\cite{Chimento:2003ta,Scherrer:2004au,Bertacca:2007ux,Bertacca:2010ct}.

In the present work, we study the proposal by Bose and Majumdar \cite{Bose:2008ew} in
which a scalar field with non canonical kinetic term accounts for the three phenomena.
This field has a quadratic potential that drives inflation and at
late times behaves almost as a purely kinetic scalar field that accounts for 
both dark matter and dark energy.  We generalize this work by introducing a  
bigger class of kinetic terms \cite{Chimento:2003ta} in which the proposal of Ref. \cite{Bose:2008ew} is a
particular case. Then, we fit the parameters using the values of the density
components in the Universe at the present time. Finally, we study the 
behavior of the field density at early times and conclude that it is possible
to obtain inflation through a potential almost in the same way as in
usual inflation models.

This work is organized as follows. In Sec. \ref{pkl} we  review  
how a purely kinetic 
Lagrangian can yield effective dark matter and dark energy densities. In Sec.
 \ref{dedm}, we  analyze the cosmological constraints to be fulfilled by the free parameters
 and in Sec. \ref{et}, we add an extra constant to our Lagrangian to build the general theory. 
 In Sec. \ref{pc}, we analyze particular cases  and show their  viability.  Then, in Sec. \ref{liper}, we study
 the effects on linear perturbations during the matter dominated era.  In Sec.  \ref{ieu},
 we analyze the behavior of the field during inflation and consider a quadratic potential 
 in slow roll.  Finally, in Sec. \ref{conclu}, we conclude.

\section{Purely kinetic Lagrangians \label{pkl}}

To start our approach, we will be working with a scalar field having a
purely kinetic Lagrangian,
\begin{equation}
	\mathcal{L} = F (X),
\end{equation}
with $X=-\frac{1}{2} \phi_{,\mu}\phi^{,\mu}$ and signature $(-,+,+,+)$. The
canonical kinetic term is recovered for $F(X)=X$.

These Lagrangians are invariant with respect to the symmetry transformation
\begin{equation}
	\phi \mapsto \phi + \phi_0.
	\label{sym}
\end{equation}
Applying the Noether theorem, there is a conserved current
$J^\mu = F_X \partial^\mu \phi$,
where $F_X$ is the derivative of the Lagrangian with respect to $X$. It
satisfies ${J^\mu}_{;\mu} = 0$, that can be transformed to
\begin{equation}
	\left( \sqrt{-g} J^\mu \right)_{,\mu} = 0.
\end{equation}
Substituting the determinant of the FRW metric, $g = -a^6$,
``$a$'' being the scale factor, and using the homogeneity property to get rid of
the spatial derivatives, one obtains the expression 
\begin{equation}
	a^6 F_X^2 X = \kappa , 
	\label{inte}
\end{equation}
with $\kappa$, an integration constant.

Another way to get this result is through the
equations of motion of these fields for a cosmological background
 \cite{Chimento:2003ta,Scherrer:2004au,Arroja:2010wy}:
\begin{eqnarray}
 3M_{\rm pl}^2 H^2 &=& \rho, \\
 \dot \rho &= & -3H(\rho + P), \label{energy}
\end{eqnarray}
where $P=F$ is the pressure,
$\rho=2XF_X-F$ the energy density,  $H=\dot a/a$ the Hubble
parameter, and $M_{\rm pl}^2 \equiv 1/(8 \pi G)$.  

Once we have specified the functional form of $F(X)$, it is possible to use the
above simplification to obtain $X$ in terms of the scale factor and then other
dynamic quantities such as the energy density and the sound velocity. In this
work, we begin with a Lagrangian proposed by Chimento in \cite{Chimento:2003ta},
\begin{equation}\label{lagrangian}
	F(X) = \frac{1}{2 \alpha - 1}\left( (AX)^\alpha
	- 2 \alpha \alpha_0 \sqrt{AX} \right) , 
\end{equation}
where the factors $A$, $\alpha$, and $\alpha_0$ are constants chosen in that
particular arrangement for the solution of the energy density to have a simple
form, as shown below. Here $A$ has dimensions of $\left[A\right]= E^{4/\alpha - 4}$,
$\alpha_0$ of $\left[\alpha_0\right] = E^{4 - 2/\alpha}$,  
where $E$ stands for energy and $\alpha$ is a dimensionless parameter. 
As the density can be expressed in terms of $X$ as $\rho =(AX)^\alpha$
using (\ref{lagrangian}), we can obtain the explicit equation of state
\begin{equation}\label{eos1}
	P = \frac{1}{2\alpha - 1} \left( \rho
	- 2\alpha \alpha_0 \rho^{1/2\alpha} \right).
\end{equation}
Also, using Eq. (\ref{inte}), it is possible to get
the energy density as a function of the scale factor
\begin{equation}\label{rhoa}
\rho = \left[ \alpha_0 + \frac{c_0}{a^3} \right]^n,
\end{equation}
with $n=2\alpha/(2\alpha-1)$ and $c_0$ an integration constant. This expression
for the density has the property that can be adjusted to behave as 
the dark densities  of the $\Lambda$CDM model, as we will show it  
in detail in the next section.

\section{Dark energy and dark matter  \label{dedm}}

There are several interesting properties of the solution (\ref{rhoa}).
It can be seen that when $a\gg \sqrt[3]{c_0 / \alpha_0}$, the density tends
to the constant value $\alpha_0^n$, and at the same time the pressure
\begin{equation}
  P = \frac{1}{(2\alpha-1)}\left[
      \left( \alpha_0 + \frac{c_0}{a^3} \right)^n
      -2\alpha \alpha_0 \left( \alpha_0 + \frac{c_0}{a^3} \right)^{n/2 \alpha}
      \right]
\end{equation}
tends to $-\alpha_0^n$, which implies that the solution approaches to a
cosmological constant. These kinds of solutions are called freezing
cosmological models \cite{Caldwell:2005tm}.  The squared adiabatic sound speed for this 
model is \cite{Garriga:1999vw}
\begin{equation} \label{sp}
   c_{s}^{2} = \frac{P_X}{\rho_X} =  \frac{1}{(2\alpha-1)} \, 
  \frac{1}{\alpha_0 a^{3} /c_{0} +1},
\end{equation}
which tends to zero for $a\gg \sqrt[3]{c_0 / \alpha_0}$. It is important for the
theories that intend to reproduce dark matter to have a small sound speed to allow the
structure formation, as we will see in detail in Sec. \ref{liper}.

Expanding the solution (\ref{rhoa}), one has that 
\begin{equation}
\rho = \sum_{k=0}^n \binom{n}{k}\alpha_0^{n-k}\left(\frac{c_0}{a^3}\right)^k,
\end{equation}
which is a finite series for $n\in \mathbb{N}$.  The first terms are 
\begin{equation} \label{expanssion1}
\rho = \alpha_0^n 
+  \underbrace{\frac{nc_0\alpha_0^{n-1}}{a^3}}_{\rho_3} 
+  \underbrace{\frac{n(n-1)c_0^2\alpha_0^{n-2}}{2a^6}}_{\rho_6}  + ... ,
\end{equation}
where the first term can be identified with the cosmological constant and the
second with the dark matter density. This can be written as
\begin{eqnarray}
   \alpha_0^n & = & \rho_{de0},\label{a}\\
  \frac{nc_0\alpha_0^{n-1}}{a_0^3} & = 
  \rho_{dm0} \approx & \frac{\rho_{de0}}{3},
   \label{b}
\end{eqnarray}
where $\rho_{de0}$ and $\rho_{dm0}$ are the current values for dark energy and
dark matter, respectively.

It has been shown that some purely kinetic models can be rewritten 
as  generalized Chaplygin gas \cite{Kamenshchik:2001cp,Bento:2002ps}, and in Ref. \cite{Sandvik:2002jz}, it 
is  shown that these models when used to unify DM and DE need to be 
indistinguishable from the $\Lambda$CDM model in order to be viable. This also   
has been proven for the affine adiabatic models in
\cite{Pietrobon:2008js,Piattella:2009kt,Bertacca:2010ct}. 
In our model, one can write $P=P(\rho)$, as shown in Eq. (\ref{eos1}), making it  similar, but 
not equal, to a generalized Chaplygin gas, since our equation of state has an extra term,    
then the above constraints do not apply here.   In our case, we 
demand  the $\rho_6$ term and subsequent in the expansion 
(\ref{expanssion1}) to be  small 
compared to the relevant terms since nucleosynthesis. In this way we avoid a different  
expansion history that would spoil the observed abundances of the light elements and 
posterior dynamics. 
This condition can be written as 
$\rho_6(a_{\rm nuc})\ll \rho_{\rm rad} (a_{\rm nuc})$. Substituting their values, we get
\begin{equation}
 \frac{n(n-1)c_0^2\alpha_0^{n-2}}{2a_{\rm nuc}^6} \ll \left(\rho_{r}\right)_0 \left(
    \frac{a_0}{a_{\rm nuc}} \right)^4,
\end{equation}
which implies
\begin{equation}\label{zc}
  z_{\rm nuc}^2 \ll
  \frac{2\left(\rho_{r}\right)_0 a_0^6}{n(n-1)c_0^2\alpha_0^{n-2}}.
\end{equation}
Using the conditions (\ref{a}) and (\ref{b}), we can obtain
\begin{equation}
	z_{\rm nuc}^2 \ll \frac{\rho_{r0}}{\rho_{de0}}  \, 36 \alpha ,
\end{equation} 
and by using the present values of the densities and the redshift at nucleosynthesis 
$z_{\rm nuc}\sim 10^{10}$, the condition becomes
\begin{equation}\label{alpha}
 \alpha \gg 10^{21} .
\end{equation}

As the parameter $\alpha$ comes from the Lagrangian (\ref{lagrangian}), the
condition (\ref{alpha}) states that the kinetic term should include the
$X$ term to a huge exponent, that is a quite unnatural election. In fact
what is happening is that the exponent in (\ref{rhoa}) is approaching to one, 
$n=1+O(10^{-21})$, in a very fine tuning manner.  Let us try to avoid this as follows.  


\subsection{Extra term \label{et}}

In the papers by Bose and Majumdar \cite{Bose:2008ew,Bose:2009kc}, it is 
considered  an additional constant to  the Lagrangian (\ref{lagrangian})  
that permits to avoid the condition (\ref{alpha}). In fact,
they choose $\alpha=1$ from the beginning and still accomplish the
observational constraints equivalent to
(\ref{a}), (\ref{b}), and (\ref{zc}).
Let us adopt this view but without fixing the value of $\alpha$. The new
Lagrangian becomes 
\begin{equation}\label{lagrangian2}
	F(X) = \frac{1}{(2\alpha-1)} \left[ (A X)^{\alpha}  
	    - 2\alpha  \alpha_0  \sqrt{AX} \right] + M.
\end{equation}

The new energy density in terms of the scale factor is
\begin{equation}\label{exp}
  \rho = \alpha_0^n - M + \frac{nc_0\alpha_0^{n-1}}{a^3} +
         \sum_{k=2}^n \binom{n}{k}\alpha_0^{n-k}\left(\frac{c_0}{a^3}\right)^k.
\end{equation}
Hereafter, we will suppose that $n\in \mathbb{N}$ therefore being the series 
finite. The conditions (\ref{a}) and (\ref{b}) are now described by
\begin{eqnarray}
   \alpha_0^n - M  &=&  \rho_{de0},\label{c} \\
   \frac{nc_0\alpha_0^{n-1}}{a_0^3}   &\approx & \frac{\rho_{de0}}{3}. 
\label{d}
\end{eqnarray}

The equivalent to (\ref{zc}) corresponds to demand the extra terms in the
energy density (\ref{exp}) to be small enough during the known evolution of the
Universe. Defining
\begin{equation}\label{rhok}
 \rho_{3k} \equiv \binom{n}{k}\alpha_0^{n-k}\left(\frac{c_0}{a^3}\right)^k,
\end{equation}
it is enough to require those terms to be small in the
nucleosynthesis epoch, i. e., $\rho_k(a_{\rm nuc})\ll\rho_r(a_{\rm nuc})$. Substituting
the values and using the condition (\ref{d}), we get
\begin{equation}
 \binom{n}{k}z_{\rm nuc}^{3k-4}\ll3^k n^k \frac{\rho_{r0}}{\rho_{de0}} 
  \left(\frac{\alpha_0^n}{\rho_{de0}}\right)^{k-1}
\end{equation}
for $k=2,3...n$. This constraint provides a condition for $\alpha_0$,  
one for each $k$, namely 
\begin{equation}\label{nk}
	\alpha_0^n \gg  \rho_{de0}\left[ \binom{n}{k}\frac{z_{\rm nuc}^{3k-4}}{(3n)^k}
	\frac{\rho_{de0}}{\rho_{r0}} \right]^{1/(k-1)} \, .	
\end{equation}
Comparing the right-hand side of the above expressions for $n$
constant and varying $k$ the most restrictive condition occurs when $k=n$. Thus, 
it is only necessary to ensure that $\alpha_0$ satisfies the inequality for
that case: 
\begin{equation}\label{nn}
	\alpha_0^{n} \gg  \rho_{de0}\left[ \frac{z_{\rm nuc}^{3n-4}}{(3 n)^n}
	\frac{\rho_{de0}}{\rho_{r0}} \right]^{1/(n-1)}.
\end{equation}

The last expression tells us that $\alpha_0^n$ has a magnitude bigger than
$10^{10}$ times the dark energy density, but from the condition (\ref{c}), the
difference between this constant and $M$ accounts for the dark energy. This
means that both constants are almost equal, with the difference needed to be fine
tuned in order to accomplish the constraints. On the other side, one advantage
of the model is that the vacuum density $M$ can be big enough to be considered
the one produced from quantum field theory considerations. This density just
needs to fulfill an expression similar to (\ref{nn}):
\begin{equation}\label{mm}
	M \gg  \rho_{de0}\left[ \frac{z_{\rm nuc}^{3n-4}}{(3 n)^n}
	\frac{\rho_{de0}}{\rho_{r0}} \right]^{1/(n-1)}.
\end{equation}
In order to avoid super-Planckian density values, we demand  
$M< M_{\rm pl}^4 \sim 10^{122} \rho_{de0}$. As the 
condition (\ref{mm}) implies at most that $M$ 
shall be bigger than $\sim 10^{29}$ times the dark energy density, we still have
$93$ orders of magnitude in which $M$ can take values. Even more because
the condition (\ref{mm}) for a particular $n$ is usually less restrictive that
the $10^{29}$ value, as we will see in the next section for specific models.

The sound speed of the model is specified in (\ref{sp}) and is negligible once $a\gg \sqrt[3]{c_0 / \alpha_0}$,
and given the constraints (\ref{d}) and   (\ref{nk}), this happens at very early times for every $n$.

\subsection{Particular cases \label{pc}}

As we mentioned in the previous subsection, the case for $n=2$, $\alpha=1$ was studied in
the papers \cite{Bose:2008ew,Bose:2009kc}. With the assumption of
$A\sim 1$, being this parameter dimensionless only in this particular case, they
found the parameter $\alpha_0$ to lie in the range
\begin{equation}\label{part1}
   10^{-48}M_{\rm pl}^2\leq 2\alpha_0 \leq 10^{-40}M_{\rm pl}^2 , 
\end{equation}
where the upper bound stems from the inflation model they used.
Assuming $\alpha_0$ to have the maximum value, the speed of sound at the
beginning of matter radiation equality turns out to be
$c_s(t_{eq})\simeq 4.1\times10^{-32}$ and  smaller thereafter.


The case with $n=3$, $\alpha=3/4$ corresponds to 
the second simplest case.  The explicit expression for the density becomes
\begin{equation} \label{rho3}
   \rho = \alpha_0^3 - M + \frac{3c_0\alpha_0^{2}}{a^3} +
          \frac{3c_0^2\alpha_0}{a^6} + \frac{c_0^3}{a^9}.
\end{equation}
From the condition for the $a^{-6}$ term to be small we obtain, as $\alpha_{0}^{3} \sim M$, that
$M \gg 0.3 z_{\rm nuc}^2 \, \rho_{de0}$ corresponding to (\ref{nk}) for $k=2$; from the
$a^{-9}$ term the condition is $M  \gg 0.7 z_{\rm nuc}^{5/2} \, \rho_{de0}$. Thus, it is 
sufficient  to demand the last condition that corresponds to $M \gg10^{25} \rho_{de0}$.

\begin{figure}
    \includegraphics[width=80mm]{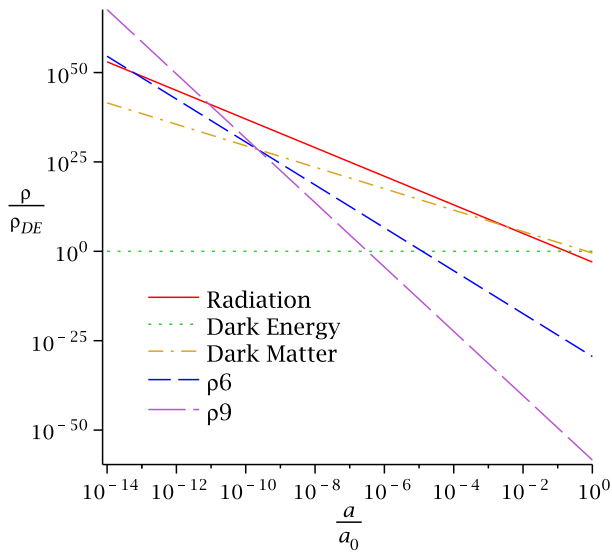}
  \caption{Evolution of the different components of the energy density for the
  $n=3$, $M=(10^{-24} M_{\rm pl})^4$ case.}
  \label{fig:evol}
\end{figure}

For example, if we set the value of $M$ to
$10^{28} \rho_{de0} \sim (10^{-24}M_{\rm pl})^4$,
the evolution of the density terms will be as shown in Fig. \ref{fig:evol},
and the redshifts at which the $\rho_6$ and $\rho_9$ terms crosses with the
radiation term are, respectively,
$z_6=1.6\times10^{14}$ and $z_9=1.3\times10^{11}$, and between them 
$z_{96}=9\times10^9$.

Having determined the value of $M$, the other parameters can be computed,  
for example, the parameter
$\alpha_0 = 2\times10^5$ eV$^{4/3}$ with the fine tuning that
$\alpha_0^3-M=(10^{-3}$eV$)^4$.
The contribution to the total Universe density of the extra terms at the moment
of nucleosynthesis and today is  $\Omega_{{\rm extra}}({\rm nuc})=10^{-6}$  and 
$\Omega_{{\rm extra}}({\rm now})=10^{-56}$,  so they 
will not affect the background dynamics  as  
compared to the $\Lambda$CDM model.

The same analysis can be carried out with any other particular election of the
parameter $n$ using (\ref{nn}) to determine $M$, and then  all the other
parameters. For example, with $n=4$, the condition will be
$M \gg 3\times10^{26}\rho_{de0}$.


\section{Linear perturbations  \label {liper}}

The models in which dark matter and dark energy come from the same scalar
field are also known as unified dark matter (UDM). The perturbation theory for UDM 
has been studied in Refs. \cite{Giannakis:2005kr,Bertacca:2010ct}, in particular,
for the purely kinetic model of Scherrer \cite{Scherrer:2004au}:
\begin{equation} \label{scherrer-model}
  F(X) = F_m + F_2(X-X_m)^2 , 
\end{equation}
where the constant $F_m$ corresponds to the negative of the dark energy density and the quadratic 
term accomplishes for a matter like component producing an energy density of the form
\begin{equation}
  \rho = -F_m + 4 F_2 X_m^2 \epsilon_1 (a/a_1)^{-3} \, ,
\end{equation}
for $X$ sufficiently close to the function minimum $X_m$, where $\epsilon = (X-X_m)/X_m$.

Our Lagrangian, given by Eq. (\ref{lagrangian2}), has a minimum since the negative term $\sqrt{AX}$  decreases as $X$ increases
for small $X$, and the term $(AX)^{\alpha}$ increases and dominates for large $X$.   The cosmological dynamics brings 
the system to the minimum, which is asymptotically reached  at 
$a =\infty$. However, one can see that the minimum is almost attained long before the 
matter domination. To see this, one can evaluate the deviation parameter $\epsilon$, that for our
model takes the form $\epsilon \approx \rho_{3}/\alpha \alpha_0^n$. The bound 
(\ref{nn}) can be used to obtain that   
\begin{equation}
  \epsilon \ll \frac{2 (n-1) \rho_{3}}{n \rho_{\rm{de}0}}
  \left( \frac{(3n)^n \rho_{r0}}{z^{3n-4}_{\rm{nuc}}\rho_{\rm{de}0}}\right)^{1/(n-1)}
\end{equation}
or
\begin{equation}\label{deviation}
   \epsilon \ll (z+1)^3 2(n-1)(3n)^{1/(n-1)}10^{(-30n+36)/(n-1)}.
\end{equation}
For the cases $n=2$ and $n=3$, the deviation at the beginning of matter domination $z_{eq}$ is smaller
than $10^{-13}$ and $10^{-16}$, respectively, and this deviation becomes even smaller with the
cosmological evolution.
Then, we can approximate our model by one of the previous type, Eq. (\ref{scherrer-model}),   
by considering a Taylor expansion around the 
minimum $X_{m} = \alpha_0^{2/(2\alpha-1)} / A$ as 
\begin{equation}
 F(X) \approx M - \alpha_0^n + \frac{1}{4} A^2 \alpha \, \alpha_0^{(2\alpha - 4)/(2\alpha - 1)}
 \left( X - X_{m} \right)^{2} \, , 
\end{equation}
which is valid long before the beginning of the matter domination, and then the
considerations on the perturbation theory for our model are the same as for Scherrer's model.

\begin{figure}
    \includegraphics[width=80mm]{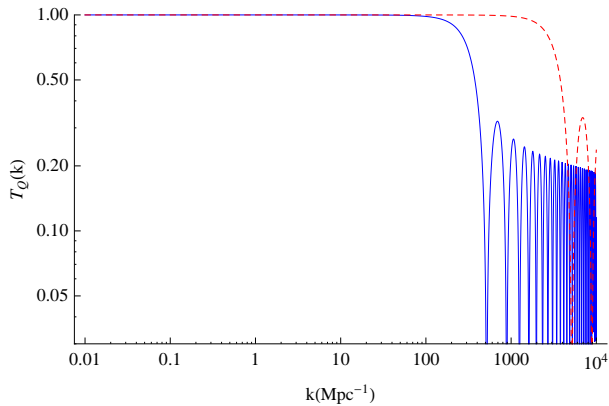}
  \caption{Deviation from the CDM transfer function for the cases $n=2$ (blue, continuos line) and $n=3$ (red, dashed line).
  The modification becomes significant only for small scales where linear theory
  is no longer valid.}
  \label{fig:trans}
\end{figure}

In the paper by Giannakis and Hu \cite{Giannakis:2005kr},
the linear perturbation theory is developed for Scherrer model and the transfer function $T(k)$
calculated. The deviation from usual cold dark matter (CDM) model can be quantified by $T_Q(k)$ defined as
$T(K) = T_Q(k) T_{\rm{CDM}}(k)$. The numerical calculation for this function can be fitted by the
function
\begin{equation}
 T_Q (k) \approx \frac{3j_1(x)}{x}\left[ 1 + (x/3.4)^2 \right]^{1/(\beta + 1)}
\end{equation}
with
\begin{eqnarray}
 x = \left( \frac{k \eta_*}{7.74} \right), & \beta = 0.21 \left[  \frac{\epsilon_0}{10^{-18}}
 \left( \frac{\Omega_m h^2}{0.14} \right)^3 \right]^{0.12},
\end{eqnarray}
where  $\eta_*$ is the conformal time evaluated at $a_*=14\epsilon_0^{1/3}$.
For $\epsilon_0\rightarrow 0$, the transfer function becomes more and more similar to the usual from CDM, i.e.,
$T_Q(k) \rightarrow 1$.  In Ref. \cite{Giannakis:2005kr}, it is concluded that its value today has to be 
smaller than $10^{-16}$ in order to yield a satisfactory CMB power spectrum. 
In our model, we can use the condition (\ref{deviation}), that for the $n=2$ and $n=3$ cases
states that the present value of the parameter $\epsilon$ is smaller than $10^{-23}$ and $10^{-26}$,  respectively, 
accomplishing a correct fit to the CMB power spectrum.   Figure \ref{fig:trans} shows the modifications to the transfer function
for these cases, where $T_Q$ deviates significantly from 1 only for small scales, which are outside the
range of validity of the linear perturbation theory. For larger $n$, the effects on $T_Q(k)$ are important for even 
larger wave numbers.

\section{Inflation and early Universe  \label{ieu}}
Now we proceed to analyze the conditions under which inflation could occur
 if the above  solution is valid at early times. Let us see how two physical 
 scales of the Universe are compared. A typical length 
 at the beginning of inflation $\lambda_i=H_i^{-1}$ will evolve as the Universe 
 expands as
\begin{equation}
	\lambda_i(a)=\lambda_i \left( \frac{a}{a_i} \right).
\end{equation}
This length should be expanded by a sufficient factor during inflation so that
at the present time it could become bigger than the length of our visible
Universe
$\lambda_0=H_0^{-1}$. The way to achieve this is by the exponential expansion
that during inflation makes $a_0/a_i = e^Na_0/a_f$, where
$a_f$ is the scale factor at the end of inflation. Usually, an expansion of
$N\approx 60$ makes the ratio
\begin{equation}
	\frac{\lambda_i(a_0)}{\lambda_0}=\frac{H_i^{-1}}{H_0^{-1}}e^N\left(
	\frac{a_0}{a_f} \right) 
	\label{long}
\end{equation}
bigger than one. If the ratio (\ref{long}) is smaller than 1, it would mean
that different parts of our Universe came from causally disconnected regions, and
we could not explain the isotropy measured, for example, in the CMB. This is
called the horizon problem.

To calculate the ratio (\ref{long}), we can use the definition
$\Omega_3 \equiv  \rho_3/3M_{\rm pl}^2H^2$ to replace the Hubble factors in the
expression. We obtain 
\begin{equation}
	\frac{\lambda_i(a_0)}{\lambda_0}= \sqrt{\frac{\Omega_{3i} \rho_{30}}
	{\Omega_{30} \rho_{3i}}} \, e^N \, \left( \frac{a_0}{a_f} \right),
\end{equation}
but as $\rho_3$ is proportional to $a^{-3}$
\begin{equation}
	\frac{\lambda_i(a_0)}{\lambda_0}= \sqrt{\frac{\Omega_{3i}}
	{\Omega_{30}} \left( \frac{a_i}{a_0} \right)} \, .
\end{equation}
In the limiting case where the amount of inflation is just enough to resolve
the horizon problem, this ratio is equal to one, in this case implying that
\begin{equation}
	\Omega_{3i} = \Omega_{30} \frac{a_0}{a_i}.
\end{equation}
But as $\Omega_{30} \approx 0.27$ and $a_0/a_i$ is typically of the order of 
$10^{51}$, the density parameter at the beginning of inflation needs to be huge.

The reason for this result is that we are demanding the dark matter density to be
present before inflation and that the expansion during this epoch does not dilute
it  too much. For that to happen as we see, the initial density needs
to be huge. To solve this problem, in the next subsection we will add a term to
the Lagrangian in order to modify the behavior of the density at early times;
essentially what we need is this term to be present only after inflation
has end. 

We could be worried  about the other terms in the expansion (\ref{exp}).
We know these terms should not be relevant at nucleosynthesis 
and thereafter, and it would be desirable to make them arbitrarily small during the inflation epoch
with the choice of a sufficiently large  $M$ parameter.  This has the effect of setting
the domination of these terms back in the preinflationary era and then
being subdominant at the beginning of inflation.  To achieve this one can constraint
the constants $\alpha_0$ and $c_0$. However, this cannot be done for the
$\rho_{3}$ term, because its value is already determined
by the observational constraints, and as we explained, it will not
be small during inflation. Thus, a new ingredient has to be added to the theory.

\subsection{Inflation driven by a potential field  \label{ikf}}

To avoid the important problems mentioned at the end of the previous 
subsection, we need to break the validity of the solution (\ref{exp}) during
inflation.  One way to achieve it is to add an interaction term,
coupled  to the inflaton that could make it to decay into the $k$-essence field
during reheating or later, and thus making
(\ref{exp}) valid only after this period. On the other side, 
the solution that we explore here is to make the $\phi$ field the only one  
responsible for the whole dynamics, including inflation.
Therefore, a potential term in the Lagrangian is necessary.

We consider the new Lagrangian
\begin{equation}\label{lagfin}
	\mathcal{L} (\phi,X) = F(X)-V(\phi),
\end{equation}
with $F$ from (\ref{lagrangian2}) and $V$ a quadratic potential
\begin{equation}
	V(\phi) =  \frac{1}{2} m^2 \phi^2.
	\label{pot}
\end{equation}
This term that will be important at early times will make the
solution to the energy density not blow up at $a\rightarrow 0$, as it happens
with (\ref{exp}).  The crucial ingredient will be considering the slow roll (sr)
approximation at which $X |_{\rm sr} \ll V |_{\rm sr} $.

The equations of motion are now
\begin{eqnarray}
	3M_{\rm pl}^2 H^2 &=& 2XF_X - F + V,  \\
	(F_X \dot \phi) \, \dot{}  + 3HF_X \dot \phi +  V' &=& 0, 
\end{eqnarray}
where $V'=dV/d\phi$.  We suppose that for the early Universe, most of the energy density is in the
potential term, so that we can use the slow roll approximation, in which the
terms in $X$ can be neglected. After inflation, the energy density will be 
transferred  to the kinetic term, so we can treat the field as purely kinetic and
proceed with the analysis made in the previous sections.

During slow roll, the equations of motion transform to
\begin{eqnarray}
	3M_{\rm pl}^2 H^2 &=& V(\phi),\label{slowrolleqn1} \\
	3HF_X \dot \phi + V'(\phi) &=& 0, 
	\label{slowrolleqn}
\end{eqnarray}
and we expect them to be valid during the inflation period, with the end of
its validity corresponding to the end of inflation.

Using the previous equations, one obtains the value 
$F_X = -V'M_{\rm pl}/\dot \phi \sqrt{3V}$, that for the case of the potential
(\ref{pot}) and $\dot \phi < 0$ can be expressed as
\begin{equation}
   F_X= \frac{mM_{\rm pl}}{\sqrt{3X}}.
   \label{X1}
\end{equation}
It is interesting to note that from this equation, the value of $X$ will
be totally defined once we have specified an expression for $F(X)$, and this
value will be approximately constant during the period of validity of the
slow roll approximation, that is, $X=X_{\rm sr}$ for all the slow roll inflation
period. This property is characteristic of the
quadratic potential chosen in (\ref{pot}); for other potentials, the right-hand
side of Eq.  (\ref{X1}) will depend on $\phi$, and the parameter $X_{\rm sr}$
no longer will be constant.

The slow roll parameters have  in this case  the form \cite{Bose:2008ew}
\begin{eqnarray}
	\epsilon &=& \frac{M_{\rm pl}^2}{2F_X} \left( \frac{V'}{V} \right)^2,
	\label{epsilon}\\
	\eta &=& \frac{M_{\rm pl}^2}{F_X^2} \frac{V''}{V}.
	\label{eta}
\end{eqnarray}
The slow roll approximation is valid as long as these parameters are small
compared to one. The only difference between these expressions and the ones with a 
canonical kinetic term is the $F_X$ factor, which is however constant
during slow roll because it depends only on  $X_{\rm sr} \approx$ const., as 
above explained.

It is considered that the end of
inflation happens when $\epsilon \sim 1$ which for the potential (\ref{pot})  
corresponds to
\begin{equation}
	\phi_f^2 = \frac{2M_{\rm pl}^2}{F_X}.
	\label{vend}
\end{equation}
If we suppose that the parameter $\eta$ also becomes one at this time 
we will have with the expression (\ref{eta}) and 
the value of the potential that $2M_{\rm pl}^{2}/ \phi_f^2 F_X^2 \sim 1$.
Substituting the value of the field at the end of inflation (\ref{vend})
leads us to $F_X(X_{\rm sr}) = 1$. Thanks to that the expressions for the slow roll
parameters (\ref{epsilon}, \ref{eta}) and the equations of motion
(\ref{slowrolleqn1}, \ref{slowrolleqn}) become the same as the canonical ones, so the subsequent
analysis can be carried out with the standard slow roll approximation.

The number of \textit{e}-folds obtained is then
\begin{equation}
	N = \int_{t_i}^{t_f} H dt = \int_{\phi_f}^{\phi_i} \frac{F_X}{M_{\rm pl}^{2}} \frac{V'}{V} dt \approx 
	\frac{(\phi_{i}^{2} - \phi_{f}^{2})}{4 M_{\rm pl}^2},
\end{equation}
where $\phi_i$ and $\phi_f$ are the field values at the beginning and at
the end of inflation, respectively.
Assuming that $N \approx 60$ and using (\ref{vend}), we obtain
\begin{equation}\label{vinf}
	\phi_i = 15.5 M_{\rm pl}.
\end{equation}
We can obtain the value of the parameter $m$ using the measured amplitude
of temperature perturbations in the CMB \cite{Smoot:1992td,Jarosik:2010iu}
$\delta = 2\times 10^{-5}$,  and as in  \cite{Lyth:1998xn,Mukhanov:2005bu}, we get
$m=7 \times 10^{-6} M_{\rm pl} (n-1)^{1/4}$.

The values for the slow roll parameters at the beginning of inflation can also
be obtained by using (\ref{vinf}) to give 
$\epsilon_i = 8.3 \times 10^{-3}$ and $\eta_i= 8.3 \times 10^{-3}$.
This justifies the use of the slow roll approximation.
Now we are in the position to calculate the spectral index $n_s=1-0.3\sqrt{n-1}$
and the tensor to scalar ratio
 $r=0.15 \sqrt{n-1}$, in accordance with the typical values in standard, chaotic 
inflation \cite{Lyth:1998xn} but with a correction due to the
noncanonical kinetic term \cite{Panotopoulos:2007ky}.

The energy scale during inflation can also be determined with the use of
Eqs. (\ref{vend}), (\ref{vinf}) and the value of $m$. The potential at the beginning and end of inflation will be, respectively,
\begin{eqnarray}
	V_i &=& 5.9 \times 10^{-9} M_{\rm pl}^4 \sqrt{n-1},  \nonumber \\
	V_f &=& 4.9 \times 10^{-11} M_{\rm pl}^4  \sqrt{n-1},
	\label{energyscale}
\end{eqnarray}
that implies that the energy scale of the potential at the beginning and end of
inflation is safely low compared to the Plank scale.


From the derivative of the second of the equations of motion (\ref{slowrolleqn})
one gets $V'' = -3 \dot H F_X$, where $\dot H$ can be obtained from the
second of the Friedmann equations 
$\dot H = - (\rho + P)/2M_{\rm pl}^2$ with $\rho + P = 2XF_X$. In this
case we arrive at $V''=3F_X^2X/M_{\rm pl}^2$. This expression can be
substituted into (\ref{eta}) to find $\eta = 3X/V$. At the end of inflation as
we supposed that $\eta$ is of order 1, we can find the value of the kinetic
parameter $X$ as being one third of the value of the potential at
that moment. This is reasonable because we think about the end of inflation as
the moment at which the potential stops its domination and gets the same order
of magnitude than the kinetic term. From (\ref{energyscale}), we find
\begin{equation}
	X_{\rm sr} \sim 1.6 \times 10^{-11}  M_{\rm pl}^4 \sqrt{n-1} .
\end{equation}


To obtain the value of $X_{\rm sr}$, we did not need the particular expression
for the kinetic term (\ref{lagrangian2}), but we have a new equation that has to
be satisfied, named $F_X(X_{\rm sr})=1$. This will be accomplished by a
suitable choice of the parameter $A$ that is the only one in the Lagrangian that
we have not constrained yet. The equation is as follows
\begin{equation}
	\frac{n}{2} \left( A^\alpha X_{\rm sr}^{\alpha-1} -
	\alpha_0 \sqrt{\frac{A}{X_{\rm sr}}} \right) = 1, 
	\label{preaprox}
\end{equation}
which cannot be exactly  solved for $A$, but we can  make
the assumption that the term containing  $\alpha_0$ is small and then obtain 
$A^{\alpha} = 2X_{\rm sr}^{1-\alpha}/n$.  For the case $n=2$, for example, we 
get $A=1$, and the neglected term in ({\ref{preaprox}}) is around $10^{-37}$ showing that 
our approximation is valid.  For $n=3$, the parameter $A$ is 
$10^{-4}M_{\rm pl}^{4/3}$ and the neglected of order $10^{-28}$.

\subsection{End of inflation}

After the end of inflation, the kinetic term will
start to dominate and the field will behave as purely kinetic with the
characteristics analyzed in the previous sections. Part of the energy density
in the field shall be transformed into radiation density that ultimately will
give rise to the electromagnetic radiation and baryonic matter present in the
Universe today. This transformation needs to be incomplete in order to let
the kinetic field behave as dark matter and dark energy. One possibility is
through gravitational particle production. Standard calculations
\cite{Ford:1986sy,Spokoiny:1993kt} indicate that this
process yields an energy density at the end of inflation given by
$\rho_{rf} \sim 0.01g H_f^2$, where the subindex refers to the end of inflation
and $g$ is the number of degrees of freedom of the particles produced that we
suppose to be $\sim 100$. Using (\ref{energyscale}), we can obtain 
$\rho_{rf}\sim 10^{-21}M_{\rm pl}^4$.

If the potential energy at the end of inflation (\ref{energyscale})
transforms into kinetic energy and starts to behave as (\ref{exp}), it will initially 
decay as $a^{-3n}$ while the radiation density 
decays only as $a^{-4}$.  Then, after an expansion of $25/(3n-4)$ \textit{e}-folds, these
densities will cross and the radiation will start to dominate.  Thus, the 
domination of the radiation term starts sooner for bigger $n$.  For example, 
for $n=2$, the Universe needs to expand $10^5$ times ($12$ \textit{e}-folds)
after the end on inflation, but for the model with $n=4$, the Universe has to
expand only $23$ times ($3$ \textit{e}-folds). After that, the model begins with the usual
radiation dominated early Universe. We can also calculate the density of
the Universe at the moment when this crossing occurs, which is given by
\begin{equation}
	\rho_c \sim V_f \left( \frac{a_f}{a_c} \right)^{3n}   \, .
\end{equation}
For example, for $n=2$, the energy will be $10^{30}$ GeV$^4$; for $n=3$, one has 
$\rho_c \sim 10^{43}$ GeV$^4$; and for bigger $n$, this energy density will be
bigger. In all the cases, the radiation dominated Universe starts well before  
the nucleosynthesis epoch which occurs at energies around 1 MeV.


\section{Conclusions  \label{conclu}}

In this work, we have generalized a particular type of solution that  
unifies  dark energy and dark matter with only one purely kinetic scalar
field. We obtained the conditions on the parameters of the Lagrangian to allow
a realistic cosmological evolution.
These conditions 
are equivalent to demand that the model be indistinguishable from $\Lambda$CDM
at least since nucleosynthesis.  
First, we noted that without the introduction of the
vacuum density $M$, the parameter $\alpha$ to which the $X$ term is raised 
in the Lagrangian had to have a value bigger than $10^{21}$, something pretty
unnatural. Once we added the $M$ parameter, we were in the position to choose
the $\alpha$ parameter.  An interesting feature of this $M$ term is that it is the only one necessary 
to cancel all the remaining $\rho_{3k}$ terms in 
(\ref{exp}), and at the same  time, it sets the proper cosmological constant today.

Even when the introduction of the vacuum density $M$ could be considered a
drawback of our model because it is equivalent to put by hand a cosmological constant,
unlike the case of introducing directly the dark energy term, 
in this case, the constant has a bigger magnitude and is easier to reconcile
with the vacuum energy obtained from quantum field theory. For example, in the
$n=3$ case, we set $M$
from around $(10^{-24} M_{\rm pl})^4$ to $M^4_{\rm pl}$. In fact, the bigger
the value of the constant term, the better this model reproduces the
$\Lambda$CDM model. Once we have chosen the value of the constant $M$, we can
calculate the other parameters in the Lagrangian in order to satisfy the
constraints.

Having demonstrated the validity of the generalized Eq. (\ref{lagrangian2})  as a 
unified dark matter and dark energy model for the background dynamics, we proceeded to 
compute its influence in the linearly perturbed Universe. In particular, we demonstrated that long 
before the matter domination era, the system is near its minimum, $\epsilon \ll 1$,  and following 
Ref. \cite{Giannakis:2005kr}, we found that transfer function of this model does not significantly deviate from 
the CDM model, emulating a perturbed, standard cosmological scenario.
 
The behavior of our solution at early times also gave us a clue that this
model should be modified in order to let the inflationary epoch to occur.
A possible solution to this scenario that had been studied in
\cite{Bose:2008ew} is to add a potential term in the Lagrangian in order to make this
field also the responsible for inflation. We analyzed this possibility in our
models and obtained the corresponding values of the remaining free parameter
in the kinetic term and the new free parameter in the potential term.
Also, it is interesting that many parameters can be fixed without specifying
the explicit expression for $F$,  but only the value of $F_X$ during inflation, 
which, for the quadratic potential in the slow roll regime is constant. It is
possible to introduce other potential terms to drive inflation following an
equivalent analysis to the one stated here.

The present work shows that the dynamics of inflation, dark energy, and dark 
matter can be unified  in a theory with a standard inflationary potential and 
a nontrivial kinetic term.   Most of the ideas present in this work were 
already stated in \cite{Bose:2008ew}, and we proved that they can be 
extended to the more general Lagrangian (\ref{lagfin}) with (\ref{lagrangian2}) 
for $n=2,3,4\cdots$.

\acknowledgments
J.L.C.C. thanks the Berkeley Center for Cosmological Physics for hospitality, and gratefully acknowledges support from a UC MEXUS-CONACYT Grant, and a CONACYT Grant No. 84133-F. J.D.S. acknowledges a CONACYT Grant No.  210405.

\bibliography{biblio}{}

\end{document}